\documentclass[aps,11pt]{revtex4}
\usepackage{epsfig}
\usepackage{amsmath}
\usepackage{bm}
\usepackage{times}

\def\beq{\begin{equation}}
\def\eeq{\end{equation}}
\def\bea{\begin{eqnarray}}
\def\eea{\end{eqnarray}}

\begin{document}
\preprint{IFIC/06-33} \vspace{5.0cm}
\title{UPPER BOUND ON THE MASS OF THE TYPE III SEESAW TRIPLET IN AN $SU(5)$ MODEL}
\author{Ilja Dor\v{s}ner$^{1}$}
\email{idorsner@phys.psu.edu}
\author{Pavel Fileviez P\'erez$^{2}$}
\email{fileviez@cftp.ist.utl.pt} \affiliation{
$^{1}$The Pennsylvania State University \\
104 Davey Lab, PMB 025, University Park, PA 16802
\\
$^{2}$Centro de F{\'\i}sica Te\'orica de Part{\'\i}culas \\
Departamento de F{\'\i}sica.\ Instituto Superior T\'ecnico \\
Avenida Rovisco Pais 1, 1049-001 Lisboa, Portugal }
\begin{abstract}
We investigate correlation between gauge coupling unification,
fermion mass spectrum, proton decay, perturbativity and
ultraviolet cutoff within an $SU(5)$ grand unified theory with
minimal scalar content and an extra adjoint representation of
fermions. We find strong correlation between the upper bound on
the mass of both the bosonic and fermionic $SU(2)$ triplets and
the cutoff. The upper bound on the mass of fermionic triplet
responsible for Type III seesaw mechanism is $10^{2.1}$\,GeV for
the Planck scale cutoff. In that case both the idea of grand
unification and nature of seesaw mechanism could be tested at
future collider experiments through the production of those
particles. Moreover, the prediction for the proton decay lifetime
is at most an order of magnitude away from the present
experimental limits. If the cutoff is lowered these predictions
change significantly. In the most general scenario, if one does
(not) neglect a freedom in the quark and lepton mixing angles, the
upper bound on the fermionic triplet mass is at $10^{5.4}$\,GeV
($10^{10}$\,GeV). Since the predictions of the model critically
depend on the presence of the higher-dimensional operators and
corresponding cutoff we address the issue of their possible origin
and also propose alternative scenarios that implement the hybrid
seesaw framework of the original proposal.
\end{abstract}
\maketitle
\section{Introduction}
The possibility to have unification of fundamental interactions is
one of the main motivations for the physics beyond the Standard
Model (SM). Partial realization of this dream is an intrinsic
feature of the so-called grand unified theories which are hence
considered the most natural extensions of the Standard Model. The
simplest grand unified theory (GUT) is the $SU(5)$ model of Georgi
and Glashow~\cite{GG}. One generation of the SM matter is
partially unified and the Higgs sector is truly minimal. This
theory is very predictive but it is certainly not realistic: one
cannot unify experimentally observed gauge couplings at the high
scale, neutrinos are massless, and unification of Yukawa couplings
of the down quarks and charged leptons at the high scale
contradicts experimentally inferred values.

Recently, there have been several efforts to define simple
realistic extensions of the Georgi-Glashow (GG) model. In
particular, it has been shown~\cite{Ilja-Pavel} that the simplest
extension with the SM matter content and an extra ${\bf 15}$
dimensional representation in the Higgs sector simultaneously
generates neutrino masses via Type II seesaw
mechanism~\cite{TypeII} and achieves unification. Different
phenomenological and cosmological aspects of this proposal have
been analyzed and reviewed in subsequent works~\cite{IPR,IPG,I,P}.
In short, this theory predicts existence of light scalar
leptoquarks and an upper bound on the proton lifetime: $\tau_p
\lesssim 1.6 \times 10^{36}$\,years. Therefore, this realistic
grand unified theory could be tested in future collider
experiments, particularly at LHC, through the production of scalar
leptoquarks and in the next generation of proton decay
experiments.

If, on the other hand, one contemplates extensions of the GG model
with extra fermions, there is another simple realistic GUT model
with an extra adjoint representation of fermions. This possibility
has been recently introduced~\cite{Borut-Goran}. The model is very
appealing since it generates two massive neutrinos via combination
of both Type III~\cite{TypeIII} and Type I~\cite{seesaw} seesaw
due to the presence of higher-dimensional operators. That model is
the primary focus of our work and we refer to it as 
``$SU(5)$ with $24_F$'' in what follows for clarity.

In this paper we study the constraints on the spectrum of the 
$SU(5)$ model proposed in~\cite{Borut-Goran} coming 
from gauge unification and perturbativity
at the one-loop level. Furthermore, we discuss in detail
correlation between the ultraviolet cutoff of the 
$SU(5)$ with $24_F$, prediction for the fermion masses and 
proton decay. We find that the upper bound on the mass of 
the fermionic $SU(2)$ triplet responsible for Type III 
seesaw mechanism depends strongly on the cutoff. The most 
exciting scenario is when the cutoff of the theory is identified 
with the Planck scale as in the original proposal~\cite{Borut-Goran}. 
In that case the discovery of those``$SU(2)$ gauginos'' is imminent. 
If that is not the case, part of the predictivity is lost in terms 
of both collider and proton decay signatures. In fact, if one lowers 
the cutoff and does (not) neglect a freedom in the quark and lepton 
mixing angles, the upper bound on the fermionic triplet mass is at $10^{5.4}$\,GeV
($10^{10}$\,GeV). Towards the end we also discuss possible origins
of the higher-dimensional operators that make the original
model~\cite{Borut-Goran} realistic and propose some alternative
scenarios.

\section{$SU(5)$ with $24_F$: Unification constraints}
The $SU(5)$ model of Georgi and Glashow~\cite{GG} is the simplest
GUT. It offers partial matter unification of one SM family $a$
($a=1,2,3$) in the anti-fundamental ${\bf \overline{5}}_{a}$ and
antisymmetric ${\bf 10}_{a}$ representations. The Higgs sector
comprises the adjoint $\bm{24}_H= (\Sigma_8, \Sigma_3,
\Sigma_{(3,2)}, \Sigma_{(\bar{3}, 2)},
\Sigma_{24})=(\bm{8},\bm{1},0)+(\bm{1},\bm{3},0)+(\bm{3},\bm{2},-5/6)
+(\overline{\bm{3}},\bm{2},5/6)+(\bm{1},\bm{1},0)$ and fundamental
$\bm{5}_H= (\Psi_D,
\Psi_T)=(\bm{1},\bm{2},1/2)+(\bm{3},\bm{1},-1/3)$ representations.
The GUT symmetry is broken down to the SM by the vacuum
expectation value (VEV) of the SM singlet in the $\bm{24}_H$
($<\bm{24}_H> \cong v/\sqrt{30} \, \textrm{diag}(2,2,2,-3,-3)$),
while the SM Higgs resides in the $\bm{5}_H$. The beauty of the
model cannot be denied. However, the model itself is not
realistic.

We are interested in the predictions of a promising extension of
the GG model with an extra fermionic adjoint $ \bm{24}_F=
(\rho_8,\rho_3, \rho_{(3,2)}, \rho_{(\bar{3}, 2)},
\rho_{24})=(\bm{8},\bm{1},0)+(\bm{1},\bm{3},0)+(\bm{3},\bm{2},-5/6)
+(\overline{\bm{3}},\bm{2},5/6)+(\bm{1},\bm{1},0) $ that has been
introduced only recently~\cite{Borut-Goran}. The nice feature of
this extension is that neutrino masses are generated using both
Type III and Type I seesaw mechanisms. Although it has been argued
that $\rho_3$ could be very light~\cite{Borut-Goran} the interplay
between perturbativity, unification constraint, fermionic mass
spectrum and proton decay has not been studied to corroborate that
claim. In this section we study this issue in order to find
correct upper bound on the mass of $\rho_3$---the field
responsible for Type III seesaw mechanism. The possibility to test
this theory through proton decay is also discussed. We also
suggest possible origins of the higher-dimensional operators that
play critical role in the original model~\cite{Borut-Goran} and
suggest alternative scenarios that implement the hybrid seesaw
framework of the original proposal.
\subsection{Gauge unification constraints}
\label{unification} In order to understand the constraints coming
from the unification of gauge couplings we use the well-known
relations~\cite{Giveon:1991zm}:
\begin{equation}
\frac{B_{23}}{B_{12}}=\frac{5}{8} \frac{\sin^2
\theta_W (M_Z) -\alpha_{em}(M_Z)/\alpha_s(M_Z)}{3/8-\sin^2 \theta_W (M_Z)}\,,
\qquad \ln
\frac{M_{GUT}}{M_Z}=\frac{16 \pi}{5} \frac{3/8-
\sin^2 \theta_W (M_Z)}{\alpha_{em}(M_Z)
 \ B_{12}}\,.
\end{equation}
where the coefficients $B_{ij}=B_i-B_j$ and $B_i = b_i+\sum_{I}
b_{iI} \ r_{I}$ are the so-called effective coefficients. Here
$b_{iI}$ are the appropriate one-loop $\beta$ coefficients of the
particle $I$ with mass $M_I$, where $r_I=(\ln M_{GUT}/M_{I})/(\ln
M_{GUT}/M_{Z})$ ($0 \leq r_I \leq 1$) is its ``running weight''.
$M_{GUT}$ is the GUT scale where the SM gauge couplings meet. We
find:
\begin{subequations}
\label{conditions}
\begin{eqnarray}
\label{condition1}
\frac{B_{23}}{B_{12}}&=&0.716 \pm0.005,\\
\label{condition2} \ln \frac{M_{GUT}}{M_Z}&=&\frac{184.9 \pm
0.2}{B_{12}}\,,
\end{eqnarray}
\end{subequations}
where we use $\sin^2 \theta_W (M_Z)=0.23120 \pm 0.00015$,
$\alpha_{em}^{-1}(M_Z)=127.906 \pm 0.019$ and
$\alpha_{s}(M_Z)=0.1176 \pm 0.002$~\cite{PDG}.

Eq.~\eqref{condition1} is sometimes referred to as the $B$-test.
It basically shows whether unification takes place or not.
Eq.~\eqref{condition2}, on the other hand, can be referred to as
the GUT scale relation since it yields the GUT scale value when
Eq.~\eqref{condition1} is satisfied. The GUT scale relation can
also bound $M_{GUT}$ for the given particle content of the theory
without any reference to Eq.~\eqref{condition1}.

The $B$-test fails badly in the SM case:
$B_{23}^{SM}/B_{12}^{SM}=0.53$. Hence the need for extra light
particles with suitable $B_{ij}$ coefficients. The $B_{ij}$
coefficients for all the particles in the GG scenario are
presented in Table~\ref{tab:table1}. Clearly, only $\Sigma_3$ can
slightly improve unification with respect to the SM case, i.e.,
$B_{23}/B_{12}=0.60$ at most. In Table~\ref{tab:table2} we shown
extra contributions to the $B_{ij}$ coefficients in the adjoint
$SU(5)$~\cite{Borut-Goran}. Notice that the field $\rho_3$ is the
only field that can further improve unification. It is thus clear
that it has to be below the GUT scale and that the upper bound on
$M_{\rho_3}$ corresponds to the smallest allowed value for
$M_{\Sigma_3}$.
\begin{table}[h]
\caption{\label{tab:table1} $B_{ij}$ coefficients in the GG
model~\cite{GG}.}
\begin{ruledtabular}
\begin{tabular}{lccccccccc}
     &Higgsless SM&$\Psi_D$&$\Psi_T$ & $V$ & $\Sigma_8$
     & $\Sigma_3$ \\
\hline $B_{23}$& $\frac{11}{3}$&$\frac{1}{6}$&$-\frac{1}{6}
r_{\Psi_T}$ &$-\frac{7}{2}r_V$ &$-\frac{1}{2}
r_{\Sigma_8}$&$\frac{1}{3} r_{\Sigma_3}$ \\
$B_{12}$&$\frac{22}{3}$&$-\frac{1}{15}$&$\frac{1}{15} r_{\Psi_T}$
&$-7r_V$ &0 &$-\frac{1}{3} r_{\Sigma_3}$ \\
\end{tabular}
\end{ruledtabular}
\end{table}
\begin{table}[h]
\caption{\label{tab:table2} Extra contributions to $B_{ij}$
coefficients in $SU(5)$ with $24_F$~\cite{Borut-Goran}.}
\begin{ruledtabular}
\begin{tabular}{lccccccccc}
     & $\rho_8$ & $\rho_3$ & $\rho_{(3,2)}$ & $\rho_{(\bar{3},2)}$ \\
\hline $B_{23}$& $- 2 r_{\rho_8}$ & $\frac{4}{3}r_{\rho_3}$ &
$\frac{1}{3} r_{\rho_{(3,2)}}$ & $\frac{1}{3} r_{\rho_{(\bar{3},2)}}$\\
$B_{12}$&  0  & $- \frac{4}{3}r_{\rho_3}$ &
$\frac{2}{3} r_{\rho_{(3,2)}}$ & $\frac{2}{3} r_{\rho_{(\bar{3},2)}}$
\end{tabular}
\end{ruledtabular}
\end{table}

Before we address the implications of the exact gauge coupling
unification within the scenario with the ${\bf 24}$ dimensional
fermionic representation we need to investigate the question of
fermion masses. Clearly, the model must rely on higher-dimensional
operators to be realistic which critically affects fermion masses
in three different sectors. Firstly, these operators correct the
GUT scale relation $Y_D=Y_E^T$, where $Y_D$ and $Y_E$ are Yukawa
matrices for the down quarks and charged leptons, respectively.
Secondly, they increase the rank of the effective neutrino mass
matrix from one to two. As a result two neutrinos are predicted to
be massive~\cite{Borut-Goran}. Thirdly, they generate mass
splitting between the fermionic fields in ${\bf 24}_F$ that is
crucial in order to achieve unification. Hence, the low-energy
predictions of the theory depend critically on the cut-off scale
that determines their maximal impact. Clearly, the higher that
scale is the more predictive the theory becomes. If the cut-off
goes to infinity one recovers renormalizable model which is not
consistent with experimental data. Thus, the cut-off scale cannot
be arbitrarily large. In the original proposal~\cite{Borut-Goran}
the cut-off ($\Lambda$) is at the Planck scale $\Lambda=M_{Pl}
\cong 1.2 \times 10^{19}$\,GeV. This yields, as the most
significant result, rather low limit on the mass of $\rho_3$ which
should be at the electroweak scale. We find that this cut-off, at
least at the one-loop level, leaves very narrow allowed region
within which both $SU(2)$ triplets---fermionic and bosonic---are
at the electroweak scale and the proton decay lifetime is an order
of magnitude below the current experimental bounds if we neglect
the quark and lepton mixings. However, we also find that if one
lowers $\Lambda$, the upper bound on the mass of $\rho_3$ relaxes
significantly.

To reach these conclusions we rely on the upper bound on the
cutoff that comes from the relation between charged fermion
masses. In this particular case the least conservative bound reads
\begin{equation}
\label{Upper} \Lambda \ \leq \ \sqrt{\frac{2}{\alpha_{GUT}}}
\times \frac{M_{GUT}} {Y_\tau - Y_b}.
\end{equation}
Eq.~\eqref{Upper} is obtained by considering the difference
between $Y_D$ and $Y_E$ at the GUT scale and assuming that the
Yukawa coefficients $Y_{ij}$ ($i,j=1,2,3$) that multiply
higher-dimensional operators remain perturbativity, i.e.,
$|Y_{ij}|\leq \sqrt{4 \pi}$. See reference~\cite{IPG} for details.
We also use the well-known $SU(5)$ relation for the masses of
proton decay mediating gauge bosons: $M_{(X,Y)}=\sqrt{5 \pi
\alpha_{GUT}/3} v$. The mass splitting between $M_X$ and $M_Y$ is
negligible for our purposes and we identified them with the GUT
scale, i.e., $M_{(X,Y)} \equiv M_{GUT}$.

Since $b$ and $\tau$ do not unify to a high degree as can happen
in supersymmetric $SU(5)$ theory the upper limit on $\Lambda$ is
determined by the mismatch between $b$ and $\tau$ masses at the
GUT scale. See for example Fig.~1 in~\cite{IPG} for behavior of
$Y_b$ and $Y_\tau$ as one goes from low energy to high energy
scales. Note that the most conservative limit on $\Lambda$ is in
fact proportional to $(Y_{\tau} + Y_b)^{-1}$. In any case, one can
find $Y_{\tau}$, $Y_b$, $\alpha_{GUT}$ and $M_{GUT}$ at any point
allowed by unification and check whether inferred $\Lambda$ from
Eq.~\eqref{Upper} is consistent with the initial assumption
$\Lambda=M_{Pl}$.

\begin{figure}[th]
\begin{center}
\includegraphics[width=4in]{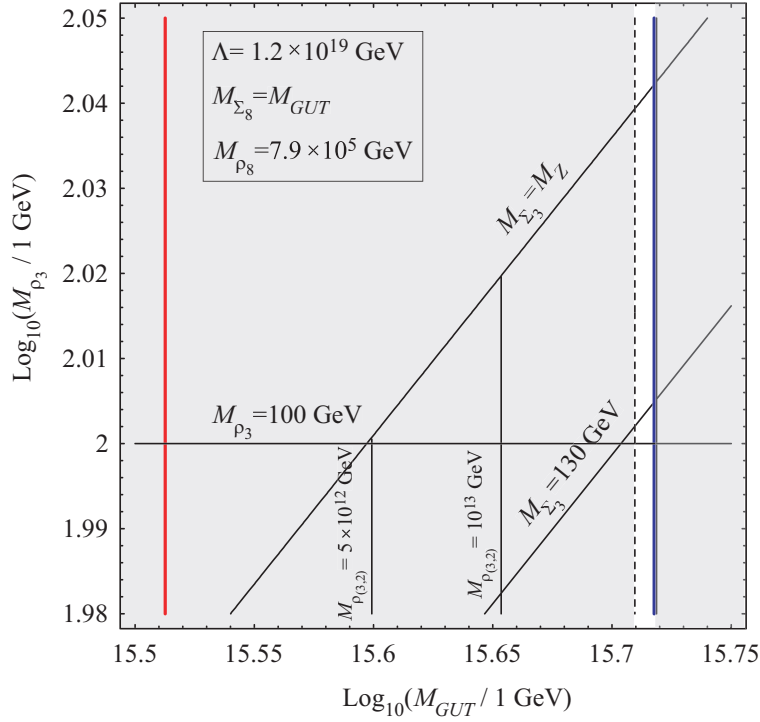}
\end{center}
\caption{\label{figure:two} Gauge coupling unification at the
one-loop level for central values of low-energy observables. Blue
line corresponds to the bound coming from perturbativity
constraints on the mass spectrum of the particles in the fermionic
adjoint, i.e.\ $a_4 \geq - \sqrt{4 \pi}$. Red line corresponds to
the up-to-date experimental bound from proton decay on $M_{GUT}$.
Dashed line corresponds to the minimal value of $M_{GUT}$ if the
cutoff is taken to be the Planck scale. In the triangle given by
the blue line, $M_{\rho_3}=100$\,GeV line and
$M_{\Sigma_3}=130$\,GeV we get $(Y_{\tau} - Y_b)=0.0038 \pm
0.0002$.}
\end{figure}

In order to find the maximal allowed value for $\Lambda$ we need
to maximize $M_{GUT}$. That turns out to be easy. Namely, there
exist a simple procedure to find maximal value for the GUT scale
as a function of $M_{\rho_8}$ as allowed by the assumption that
Planck scale effects split multiplets of ${\bf 24}_F$. We will
explain all the details later. What is important at this point is
the following. If we set $M_{\rho_3}=M_{\Sigma_3}=M_Z$ and
$M_{\Sigma_{8}}=M_{GUT}$ we get $M_{GUT} \cong 10^{15.74}$\,GeV
for $\alpha_{GUT}^{-1} \cong 35.4$, $M_{\rho_{(3,2)}}=2.5 \times
10^{13}$\,GeV and $M_{\rho_{8}}=6.3 \times 10^{5}$\,GeV at the
one-loop level. At the same time we find $(Y_{\tau} - Y_b)=0.0038
\pm 0.0002$ where the errors are associated to the $1 \sigma$
variation of the $b$ quark mass at the $M_Z$ scale. We take as
input $m_b=2.89 \pm 0.11$\,GeV and
$m_\tau=1.74646^{+0.00029}_{-0.00026}$\,GeV at $M_Z$~\cite{IPG}.
This in turn implies via Eq.~\eqref{Upper} that $\Lambda$ barely
reaches the Planck scale ($\Lambda=1.22 \times
10^{19}\,\textrm{GeV}=M_{Pl}$). Since $(Y_{\tau} - Y_b)$ remains
almost constant in the region of interest and $M_{GUT}$ can only
be below its maximal value that means that if the cutoff of the
theory is the Planck scale the allowed parameter space is
extremely narrow. To illustrate our point we plot the allowed
parameter space for $M_{\Sigma_3}$ and $M_{\rho_{(3,2)}}$ in
$M_{GUT}$--$M_{\rho_3}$ plane in Fig.~\ref{figure:two}. (Note,
whenever we refer to $\rho_{(3,2)}$ we also refer to
$\rho_{(\bar{3},2)}$ since
$M_{\rho_{(3,2)}}=M_{\rho_{(\bar{3},2)}}$.) The region to the left
of the dashed line in Fig.~\ref{figure:two} is excluded through
the use of Eq.~\eqref{Upper}. More precisely, the dashed line is
obtained by setting $\Lambda=M_{Pl}$ and then plotting the
smallest possible value of $M_{GUT}$ as allowed by
Eq.~\eqref{Upper}. The region to the right of the blue line is
eliminated by the perturbativity constraints on the spectrum of
fermionic particles in the adjoint representation as we discuss
later. The only viable parameter space is the strip between the
blue and dashed line. This scenario could clearly be tested at the
LHC since both the bosonic and fermionic triplets are light.
Moreover, proton decay is only factor 3--6 away from the current
bound on the proton lifetime. To move the strip to the right one
needs to either lower slightly mass of $\Sigma_8$ or $\rho_8$.
This however would make the allowed region disappear once $M_{GUT}
\cong 10^{15.74}$\,GeV is reached. Since the prediction for
$M_{\rho_3}$ is practically at the present experimental limit on
its mass $M_{\rho_3}>100$\,GeV~\cite{PDG}, which admittedly is
model dependent, the two-loop analysis would probably be in order.
We do not attempt that since even the full one-loop treatment
would also have to include influence of higher-dimensional
operators on the gauge coupling unification
conditions~\cite{Hill:1983xh,Shafi:1983gz} which we neglected.

Unfortunately, we do not know what the cutoff(s) is (are) and we
also have to see how the predictions of $SU(5)$ with $24_F$ hold if we
allow $\Lambda$ to vary within reasonable range $10 M_{GUT} \leq
\Lambda \leq M_{Pl}$.

\begin{figure}[th]
\begin{center}
\includegraphics[width=4in]{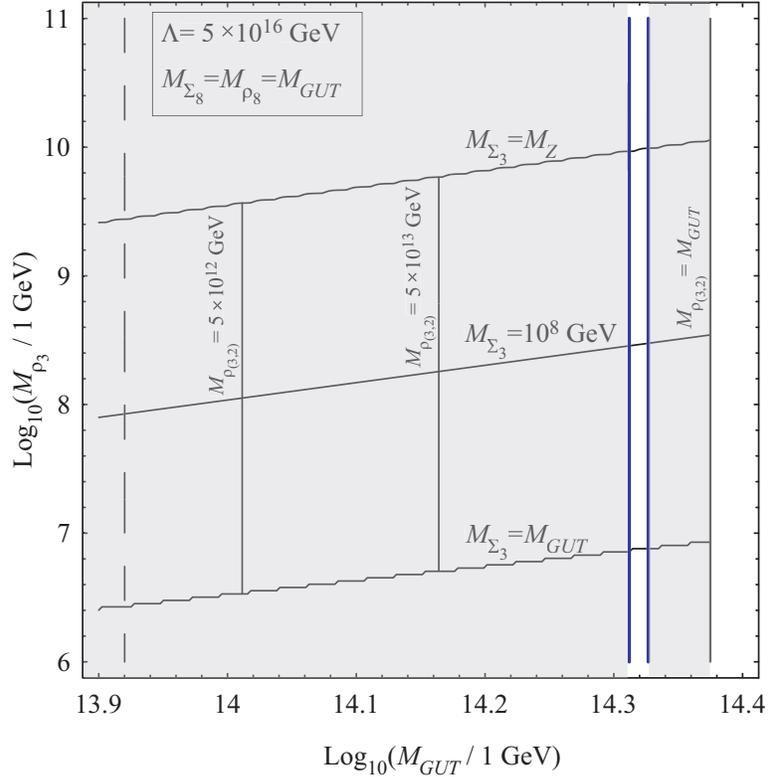}
\end{center}
\caption{\label{figure:one} The gauge coupling unification at the
one-loop level for central values of low-energy observables. Blue
lines correspond to the bounds coming from perturbativity, i.e.
$|a_4| \leq \sqrt{4 \pi}$. The dashed line in is the experimental
bound on $M_{GUT}$ if $d=6$ gauge mediated proton decay is
suppressed.}
\end{figure}

Before we answer what happens if we lower $\Lambda$ let us discuss
the correct procedure to obtain exact gauge coupling unification
within this scenario in view of the fact that the masses of
particles within the fermionic adjoint are related to each other.
The masses of relevant particles in ${\bf 24}_F$ to the leading
order in $1/\Lambda$ are~\cite{Borut-Goran}:
\begin{eqnarray}
M_{\rho_0} &=& m_F -\frac{\lambda_F v}{\sqrt{30}}+
\frac{v^2}{\Lambda} \left[a_1+a_2+\frac{7}{30}(a_3+a_4) \right],\\
M_{\rho_3} &=& m_F -\frac{3 \lambda_F v}{\sqrt{30}}+
\frac{v^2}{\Lambda} \left[a_1+\frac{3}{10}(a_3+a_4)\right],\\
M_{\rho_8} &=& m_F +\frac{2 \lambda_F v}{\sqrt{30}}+
\frac{v^2}{\Lambda} \left[a_1+\frac{2}{15}(a_3+a_4)\right],\\
M_{\rho_{(3,2)}} &=& m_F -\frac{\lambda_F v}{2 \sqrt{30}}+
\frac{v^2}{\Lambda} \left[a_1+\frac{(13 a_3- 12 a_4)}{60}\right].
\end{eqnarray}
It is understood that $\lambda_F$ and $a_i$ ($i=1,2,3$) should be
perturbative. We hence demand that $|a_i| \leq \sqrt{4 \pi}$. In
that case we obtain
\begin{equation}
\label{a4} a_4=\frac{2 \ \Lambda \ \pi \ \alpha_{GUT} \left(
(M_{\rho_8} - 2 M_{\rho_{(3,2)}})+M_{\rho_3} \right)}{M_{GUT}^2},
\end{equation}
where we again use $M_{(X,Y)}=\sqrt{5 \pi \alpha_{GUT}/3}
v=M_{GUT}$. Clearly, $M_{\rho_8}$, $M_{\rho_{(3,2)}}$ and
$M_{\rho_3}$ are not independent from each other. In fact,
perturbativity of $a_4$ implies that there are three regimes:

\begin{itemize}

\item In the first regime, $M_{\rho_8}$ and $M_{\rho_{(3,2)}}$
could both be at the GUT scale as long as $M_{\rho_8} \cong 2
M_{\rho_{(3,2)}}$. In that case $M_{\rho_3}$ must be of the order
of $M_{GUT}^2/\Lambda$ or smaller where it wants to be anyway in
order to fix the B-test. In this regime the GUT scale is too low
to be in agreement with the experimental limits on proton decay
lifetime unless one allows for special realization of fermion mass
matrices~\cite{upper}. Even though we do not advocate this
scenario it is important to realize that this is still allowed by
experimental data and the structure of the model. In that case one
needs $\Lambda$ to be around $10^{17}$\,GeV in order to suppress
proton decay by adjusting Yukawa couplings of ordinary
matter~\cite{IPG}. We show one such example in
Fig.~\ref{figure:one}.

\item In the second regime, $M_{\rho_3}$ is small and $M_{\rho_8}$
and $M_{\rho_{(3,2)}}$ are both of the order of
$M_{GUT}^2/\Lambda$. The cancellation between the two masses does
not have to be as efficient as in the first case and one can
easily see that the upper bound on $M_{GUT}$ comes when
$M_{\rho_8}<M_{\rho_{(3,2)}}\sim M_{GUT}^2/M_{Pl}$, i.e., $a_4 = -
\sqrt{4 \pi}$. Recall, light $\rho_{(3,2)}$ spoils unification.
Hence, in the large $M_{GUT}$ regime $\rho_{(3,2)}$ has to be as
heavy as possible. This scenario is promising since large GUT
scale implies light $\rho_3$. It is this scenario that is
advocated in~\cite{Borut-Goran}. Example given in
Fig.~\ref{figure:two} corresponds to this scenario once we fix
$\Lambda=M_{Pl}$.

\item There is however a third scenario that interpolates between
the two. The nice feature of this scenario is that it yields
correct upper bound on the mass of $\rho_3$. It can be best
grasped by considering the case when both $M_{\rho_8}$ and
$M_{\rho_3}$ are below the GUT scale. In this case the GUT scale
is guaranteed to be large. It is then easy to see that if
$M_{\rho_{(3,2)}}=M_{GUT}$ and $a_4\geq-\sqrt{4 \pi}$ one obtains
\begin{equation}
\label{new} \Lambda\leq \frac{M_{GUT}}{ 2 \alpha_{GUT}
\sqrt{\pi}}.
\end{equation}
We will next discuss this case to establish an upper bound on
$M_{\rho_{3}}$.

\end{itemize}

Clearly, since in this theory $\alpha_{GUT}^{-1}\sim 35$ and
proton decay experiments require $M_{GUT}> (2$--$3) \times
10^{15}$\,GeV then Eq.~\eqref{new} implies that we need a scenario
where $\Lambda\simeq 10 M_{GUT}\sim 2$--$3 \times 10^{16}$\,GeV.
This then allows us to maximize $M_{\rho_3}$ in order to
understand the testability of the model. So, to maximize
$M_{\rho_3}$ at the one-loop level is simple. All one needs is to
set $M_{\rho_{(3,2)}}=M_{\Sigma_8}=M_{GUT}$, $M_{\Sigma_3}=M_Z$
and solve for $M_{\rho_8}$, $M_{\rho_3}$ and $M_{GUT}$ using
B-test, the GUT scale relation and Eq.~\eqref{Upper} where
$M_{GUT}$ must match the most stringent constraint coming from
proton decay. We find that $M_{\rho_3}=2.3 \times 10^5$\,GeV,
$M_{\rho_8}=2.0 \times 10^9$\,GeV, $M_{GUT}=3.1 \times
10^{15}$\,GeV, $\alpha_{GUT}^{-1}=38.7$ and accordingly
$\Lambda\leq 3.4 \times 10^{16}$\,GeV. So, the correct upper bound
on $M_{\rho_3}$ at the one-loop level is $M_{\rho_3}=2.3 \times
10^5$\,GeV. One can also check that $v/\Lambda\geq0.12$ which
implies that higher order corrections in $1/\Lambda$ on
Eq.~\eqref{Upper} cannot significantly affect our conclusions.

As far as experimental proton decay constraints are concerned we
match $M_{GUT}$ with the experimental limit from the dominant
proton decay mode in $SU(5)$ which is usually taken to be $p
\rightarrow \pi^0 e^+$. The theoretical prediction for this
channel can be summarized in the following way: $\tau^{tho}=1.2
\times 10^{32} (M_{GUT}/10^{16}\,\textrm{GeV})^4 \alpha_{GUT}^2
(\alpha/0.015\,\textrm{GeV}^3)$\,years. Here, $\alpha$ is a
relevant matrix element. So, the current experimental limit
$\tau>1.6 \times 10^{33}$\,years~\cite{PDG} translates into the
following bound on $M_{GUT}$: $M_{GUT}>(1.6/1.2)^{1/4} 10^{16}
\sqrt{\alpha_{GUT}^{-1}}$\,GeV. Red line in Fig.~\ref{figure:two}
is generated using this result.

Let us finally address the case when we allow for suppression of
the proton decay through gauge boson mediation~\cite{upper} in the
adjoint $SU(5)$ model. That case would correspond to the first
scenario when $M_{\rho_8} - 2 M_{\rho_{(3,2)}} \simeq 0$. So,
allowed region would be very narrow strip given by the allowed
range for $a_4$. We show one such scenario in
Fig.~\ref{figure:one}, where the vertical blue lines correspond to
the bounds coming from perturbativity, i.e. $|a_4| \leq \sqrt{4
\pi}$. From this plot we can basically find the upper bound on the
mass of the fermionic $SU(2)$ triplet $\rho_3$ responsible for the
Type III seesaw mechanism to be $M_{\rho_3} \lesssim
10^{10}$\,GeV. This bound clearly reflects the worst case scenario
as far as the testability of the model is concerned. The dashed
line in Fig.~\ref{figure:one} is the experimental bound on
$M_{GUT}$ if $d=6$ gauge mediated proton decay is suppressed. It
corresponds to $M_{GUT} > 5.1 \times 10^{14}
\sqrt{\alpha_{GUT}}$\,GeV~\cite{upper}.

We stress again that the maximal value of the GUT scale depends
crucially on the cutoff of the theory. In fact, the highest value
that the GUT scale can reach at the one-loop level in this model
is basically $M_{GUT}=2 \times 10^{16}$\,GeV which corresponds to
setting $B_{12}^{(min)}=28/5$ in Eq.~\eqref{condition2}. With that
in mind we make the following two comments regarding proton decay.
If we neglect the fermionic mixings we can make a naive estimation
of the upper bound on the proton decay lifetime to be $\tau_p
\lesssim 10^{36-37}$ years. Of course, the absolute upper bound,
if we use the whole freedom in the Yukawa sector, reads as
$\tau_p^{24_F} \ \lesssim \ 5.2 \ \times \ 10^{42}\,\text{years}$,
where we set $\alpha_{GUT}=1/37$ and $\alpha=0.015$ GeV$^3$ for
the matrix element.

\subsection{The origin of higher-dimensional operators and alternative scenarios}

Before we conclude, let us address the issue of the possible
origin of higher-dimensional operators. After all, they play
decisive role in making the model $SU(5)$ with $24_F$ realistic.

The original model~\cite{Borut-Goran} includes higher-dimensional
operators that are suppressed by the Planck scale. We have
confirmed that these operators are indeed sufficient, albeit
barely, to make the model viable. Operators of this sort are
expected to appear on general grounds as harbingers of
gravitational physics where the only relevant scale is the Planck
scale. As such, they have been extensively used in the grand
unified model building ever since they were proposed to correct
the Georgi-Glashow fermion mass predictions~\cite{Ellis:1979fg}.

If, however, the cutoff scale is below the Planck scale one might
ask for the possible renormalizable model that would effectively
mimic the original model. To this end we observe that one does not
need to invoke nonrenormalizable operators at all nor any
``exotic'' physical setup. Namely, in order to have the most
minimal renormalizable setup that yields the original model, it is
sufficient to introduce the following additional matter
representations: $\bm{1}$, $\bm{5}$, $\overline{\bm{5}}$ and
$\bm{24}$. These can clearly have masses above the GUT scale that
can be identified with the scale $\Lambda$. Once these fields are
integrated out the effective model would have exactly the same
features as the original model~\cite{Borut-Goran}. In particular,
the established upper limits on the masses of $\rho_3$ and
$\Sigma_3$ would still hold as well as the upper bound on
$M_{GUT}$.

There is another rather different renormalizable realization of
the model in question. Namely, if one introduces an additional
$\bm{45}$-dimensional Higgs field one can simultaneously generate
both the charged~\cite{Georgi:1979df} and neutrino masses at the
renormalizable level~\cite{Pavel-45}. However, in this case the
predictions are quite different. This is primarily due to the fact
that there are more fields that can potentially contribute to the
running of the gauge couplings and proton decay. Moreover, the
possible mass spectrum of the fields in $24_F$ is rather
different~\cite{Pavel-45}.

The $SU(5)$ scenario that incorporates combination of Type I and
Type III seesaw due to the presence of extra fermionic adjoint is
tailor-made for applications within the extra-dimensional setup.
Here, in particular, we have in mind a five-dimensional
nonsupersymmetric framework~\cite{Kawamura:2000ir,Haba:2001ci}. In
this approach the $SU(5)$ symmetry of the five-dimensional bulk is
reduced to the effective SM symmetry on one four-dimensional brane
and $SU(5)$ symmetry on the other brane by compactification upon
$S^1/(Z_2 \times Z'_2)$. Such a setup would naturally accommodate
doublet-triplet splitting if the SM Higgs field originates from
the bulk. Also, since the symmetry breaking would be accomplish
using the judicious parity assignment under $Z_2$ and $Z'_2$ there
would not be any need for the adjoint Higgs representation. In
other words, there is a possibility to have a rather predictive
setup due to potentially very small number of light extra fields
with respect to the SM particle content. In addition, it would be
possible to completely suppress proton decay for particular
locations of matter fields. One particular nice feature of this
framework is that the parity assignment that yields the SM on one
of the branes generates same parity properties for
$(\bm{8},\bm{1},0)$, $(\bm{1},\bm{3},0)$ and $(\bm{1},\bm{1},0)$
in the fermionic adjoint. This would guarantee, unlike in the
ordinary four-dimensional framework, that the least massive
fermionic triplet and singlet states are degenerate as long as
they originate from the bulk. In addition, this assignment
automatically insures that no anomalies are introduced at the
branes. One possible scenario would be to have the $\bm{5}$ and
$\bm{45}$ dimensional Higgs representations in the bulk along with
the gauge fields. In that case it would be possible to build the
model with all the matter fields located on the $SU(5)$ brane.
\section{Summary and discussions}
We investigated the relation between perturbativity, unification
constraint, prediction for fermion masses, proton decay and
ultra-violet cutoff $\Lambda$ within the $SU(5)$ grand unified
theory with minimal scalar content and an extra adjoint
representation of fermions. If the cutoff is at the Planck scale
the upper bound on the mass of the Type III triplets is
practically at the current experimental limit $M_{\rho_3} <
10^{2.1}$\,GeV. In that case both the idea of grand unification
and nature of seesaw mechanism could be tested at collider
experiments through the production of those particles. Moreover,
the prediction for the proton decay lifetime is at most an order
of magnitude away from the present experimental limits. If the
cutoff is below the Planck scale we find that the upper bound on
the mass of the fermionic $SU(2)$ triplet responsible for Type III
seesaw mechanism is $M_{\rho_3} \lesssim 2 \times 10^{5}$\,GeV if
we use the strongest constraints on the GUT scale coming from
proton decay, $M_{GUT} > (2$--$3) \times 10^{15}$\,GeV. Finally,
if we allow for suppression of proton decay operators using the
full freedom of the model the limit is not relevant for collider
physics at all and it reads $M_{\rho_3} < 10^{10}$\,GeV. Since the
predictions of the model depend critically on the cutoff we have
addressed the issue of the possible origin of the
higher-dimensional operators and proposed some alternative
scenarios.

Let us finally compare our results with the results presented in
Refs.~\cite{Borut-Goran,Borut-Goran2}. Firstly, we show that
$M_{Pl}$ can be the UV cutoff of $SU(5)$ with $24_F$. This is in
conflict with the results presented in Ref.~\cite{Borut-Goran2}
where the authors retract their initial claim~\cite{Borut-Goran}.
Secondly, the upper bound on the mass of
$\rho_{(3,2)}$---$M_{\rho_{(3,2)}}  <  M_{GUT}^2 / \Lambda$---as
suggested in Refs.~\cite{Borut-Goran,Borut-Goran2} depends on the
specific assumptions about the cutoff and hence does not reflect
the full parameter freedom of the model. See in particular
Eq.~(11) in~\cite{Borut-Goran} and Eq.~(12)
in~\cite{Borut-Goran2}. In fact, we show that $\rho_{(3,2)}$ could
be at the GUT scale. As a consequence, we obtain the upper bound
on the mass of $\rho_3$ that is two orders of magnitude above the
bounds suggested in Refs.~\cite{Borut-Goran,Borut-Goran2} if we
neglect the quark and lepton mixing angles. Thirdly, we show that
the absolute upper bound on the mass of $\rho_3$, the field
responsible for Type III seesaw, is $10^{10}$\,GeV if we use the
full freedom of the model. This freedom has not been accounted for
elsewhere.

\begin{acknowledgments}
We would like to thank German Rodrigo for discussion and
collaboration during early stages of this work. We are grateful to
Goran Senjanovi\'{c} on illuminating discussion and shared
insight. The work of P.~F.~P.\ has been supported by {\em
Funda\c{c}\~{a}o para a Ci\^{e}ncia e a Tecnologia} (FCT,
Portugal) through the project CFTP, POCTI-SFA-2-777 and a
fellowship under project POCTI/FNU/44409/2002.
\end{acknowledgments}

\end{document}